\documentclass[aps,prd,twocolumn,superscriptaddress,preprintnumbers,floatfix,nofootinbib,notitlepage,showkeys,showpacs]{revtex4-1}
\pdfoutput=1

\usepackage[utf8]{inputenc}
\usepackage{cancel}

\usepackage{graphicx}
\usepackage{hyperref}
\usepackage{latexsym}
\usepackage{amsmath}
\usepackage{amssymb}
\usepackage{bbm}

\usepackage{ulem}
\usepackage{pdfsync}
\usepackage{epsfig}
\usepackage{epstopdf}
\usepackage{subfigure}
\usepackage{color}
\usepackage{comment}
\usepackage{slashed}
\usepackage{cancel}

\def\beq{\begin{equation}}
\def\eeq{\end{equation}}
\def\baq{\begin{eqnarray}}
\def\eaq{\end{eqnarray}}

\newcommand{\be}{\begin{equation}} 
\newcommand{\ee}{\end{equation}}
\newcommand{\bea}{\begin{eqnarray}} 
\newcommand{\eea}{\end{eqnarray}}

\newcommand{\bmp}{\noindent\begin{minipage}{16cm}}
\newcommand{\emp}{\end{minipage}\vskip 7mm} 
\def\lsim{\mathrel{\raise.3ex\hbox{$<$\kern-.75em\lower1ex\hbox{$\sim$}}}}
\def\gsim{\mathrel{\raise.3ex\hbox{$>$\kern-.75em\lower1ex\hbox{$\sim$}}}}

\newcommand{\intron}[1]{}

    \newcommand{\SO}{\mathrm{SO}}
    
    \newcommand{\SUL}{\mathrm{SU}(2)_{\mathrm{L}}}

\begin{document}

\title{Inflation and pseudo-Goldstone Higgs boson
}

\author{Tommi Alanne}
\email{alanne@cp3.sdu.dk}
\affiliation{{CP}$^{ \bf 3}${-Origins}, 
University of Southern Denmark, Campusvej 55, DK-5230 Odense M, Denmark}
\author{Francesco Sannino}
\email{sannino@cp3.dias.sdu.dk}
\affiliation{{CP}$^{ \bf 3}${-Origins},  University of Southern Denmark, Campusvej 55, DK-5230 Odense M, Denmark}
\affiliation{The Danish Institute for Advanced Study {\rm{Danish IAS}},  University of Southern Denmark, Campusvej 55, DK-5230 Odense M, Denmark}
\author{Tommi Tenkanen}
\email{t.tenkanen@qmul.ac.uk}
\affiliation{Department of Physics, University of Helsinki \& Helsinki Institute of Physics, \\
                      P.O.~Box 64, FI-00014, Helsinki, Finland}
                      \affiliation{Astronomy Unit, Queen Mary University of London, \\
Mile End Road, London, E1 4NS, U.K.}
\author{Kimmo Tuominen}
\email{kimmo.i.tuominen@helsinki.fi}
\affiliation{Department of Physics, University of Helsinki \& Helsinki Institute of Physics, \\
                      P.O.~Box 64, FI-00014, Helsinki, Finland}

\begin{abstract}
We consider inflation within a model framework where the Higgs boson arises as a pseudo-Goldstone boson
associated with the breaking of a global symmetry at a scale significantly larger than the electroweak one.
We show that in such a model the scalar self-couplings can be parametrically suppressed
and, consequently, the non-minimal couplings to gravity can be of order one or less,
while the inflationary predictions of the model remain compatible with the precision cosmological observations.
Furthermore, in the model we study, the existence of the electroweak scale is entirely due
to the inflaton field. Our model therefore suggests that inflation and low energy particle phenomenology
may be more entwined than assumed so far.
\end{abstract}

\preprint{CP$^3$-Origins-2016-049 DNRF90 \& HIP-2016-32/TH}

%
\maketitle

%
\section{Introduction}

Cosmic inflation and electroweak (EW) physics may be profoundly connected. The well-known
concrete realization of this connection is the Higgs inflation model~\cite{Bezrukov:2007ep}  (for early studies on the topic, see for example \cite{Spokoiny:1984bd,Salopek:1988qh,PhysRevD.39.399,CervantesCota:1995tz}), which is favoured by its simplicity and
conformity with observations~\cite{Ade:2015lrj}. At classical level, it provides
clear predictions for inflationary parameters and can, in principle, be used to accurately calculate the subsequent evolution of the universe including the stage of
reheating~\cite{GarciaBellido:2008ab,Bezrukov:2008ut}.

However, the Standard Model (SM) Higgs inflation suffers from a few unavoidable
theoretical disadvantages: the Higgs field, $H$, must couple to gravity, $\xi |H|^2 R$,
with a very large non-minimal coupling, $\xi=\mathcal{O}(10^4)$, to
sufficiently flatten the potential at high field values; the renormalization group (RG) running drives the quartic self-coupling $\lambda_{H}$ of the Higgs negative at scales well below the inflationary scale\footnote{A detailed analysis unveiling and abiding the Weyl consistency conditions has been performed in \cite{Antipin:2013sga}.}~\cite{Degrassi:2012ry}; and there have been raised concerns about possible unitarity violation at scales below the inflationary one \cite{Burgess:2009ea,Barbon:2009ya,Barvinsky:2009ii,Bezrukov:2010jz,Burgess:2010zq,Hertzberg:2010dc} (see, however, e.g. \cite{Calmet:2013hia,Burgess:2014lza,Bezrukov:2014ipa,Fumagalli:2016lls,Enckell:2016xse} for further discussion).

In addition to the Higgs, one can include additional scalar fields singlet under
all SM symmetries and non-minimally coupled to gravity \cite{Salopek:1988qh, Lerner:2009xg,Okada:2010jf,GarciaBellido:2011de,Bezrukov:2013fca,Barbon:2015fla,Tenkanen:2016idg, Marzola:2016xgb,Tenkanen:2016twd}, or consider alternative non-minimal
couplings~\cite{Germani:2010gm,Nakayama:2010sk,Kamada:2010qe,Kamada:2012se, vandeBruck:2015gjd}.
The model where the inflaton is identified with a non-minimally coupled singlet scalar $S$ suffers from the same problem of a large non-minimal coupling, $\xi_S = 49 000\sqrt{\lambda_S}$. This problem
of $S$ inflation can be alleviated if a sufficient hierarchy
$\lambda_S\ll  \lambda_H$ exists, and in~\cite{Okada:2010jf,Bezrukov:2013fca,Tenkanen:2016twd} it was shown that inflation can be realized with $\xi_S=\mathcal{O}(1)$ or less if $\lambda_S\lesssim 10^{-8}$. Therefore, it would be desirable to identify model frameworks where small scalar self-interactions are generated and stability of the scalar potential can be ensured up to the inflationary scale.

One possible model framework, called {\it elementary Goldstone Higgs} (EGH),
where these goals can be achieved, has been introduced
in~\cite{Alanne:2014kea}. This model is based on an elementary scalar sector with
a global symmetry larger than in the SM, and this symmetry is explicitly
broken by the coupling with the EW gauge currents and the SM Yukawa interactions.
Under radiative corrections, these sources of explicit breaking will align the vacuum
with respect to the EW symmetry. As shown in~\cite{Alanne:2014kea}, the
vacuum aligns very near the vacuum where the EW symmetry remains unbroken, and consequently the
125-GeV Higgs boson is identified with a pseudo-Goldstone boson (pGB) of the breaking of the global symmetry.

The EW scale in this model framework is therefore induced by physics operating
at some possibly much higher scale.
In~\cite{Alanne:2015fqh} it was shown that the EGH model framework, within a Pati--Salam-type unification scenario,
can be applied to explain the large hierarchy between the EW and unification scales.
As a consequence of this large hierarchy, all scalar self-couplings become parametrically small.

In this paper we consider inflation in this model context, where the Higgs boson arises as a pGB. In Sec.~\ref{model},
we introduce a minimal model where this dynamics can be realized. In Sec.~\ref{inflation}, we consider the inflationary
dynamics. Our main new results are that non-minimal coupling with gravity can be
small, of order one or less, and the resulting inflationary dynamics is consistent with current observations.
Our results also illustrate how the inflaton field intertwines with the low-energy
particle phenomenology: in the model we consider, the existence of the EW scale is due to
the inflaton field itself. In Sec.~\ref{vacuumstability}, we shortly discuss the stability of the EW vacuum against fluctuations both during and after inflation. Finally, in Sec.~\ref{checkout}, we conclude and present outlook for further work.

\section{The minimal setup}
\label{model}

It was shown in~\cite{Alanne:2016mmn} that the minimal model able to incorporate an elementary pGB Higgs boson by enlarging only the Standard Model scalar sector features an $\SO(5)\rightarrow\SO(4)$ global symmetry breaking pattern, and contains
an additional singlet scalar. The scalar sector can be conveniently parametrized by fields $\Sigma$ describing a linear
$\sigma$ model based on the coset $\SO(5)/\SO(4)$, and a real singlet, $S$. 
As explained in detail in~\cite{Alanne:2016mmn}, the introduction of the singlet $S$ ensures that the EW symmetry is properly broken in the vacuum.
The EW symmetry, $\SUL\times \mathrm{U}(1)_Y$, is embedded in $\SO(5)$, and the field $\Sigma$ contains an EW Higgs doublet, $H$, and another
real singlet, $\varphi$. The radiative symmetry breaking dynamics imply $\langle H^2\rangle=v^2\sin^2\theta\equiv v_{\rm{w}}^2$ and $\langle \varphi^2\rangle=v^2\cos^2\theta$. The $\theta$ angle parametrizes the compact flat direction associated to the freedom in choosing the vacuum orientation at tree-level. Radiative corrections single out a value for $\theta$ and the large hierarchy $v_{\rm{w}}\ll v$ is reflected in $\sin\theta\ll 1$. 

The scalar potential for these fields, assuming a $Z_2$-symmetric singlet sector for simplicity, is given by
    \begin{equation}
	\label{V0}
	\begin{split}
	    V_0=&\,m_H^2 H^{\dagger}H+\frac{1}{2}m_{\varphi}^2 \varphi^2+\frac{1}{2}m_S^2S^2\\
	    &+ \lambda_H(H^{\dagger}H)^2+\frac{\lambda_{\varphi}}{4}\varphi^4+\lambda_{HS}H^{\dagger}H\, \varphi^2\\
	    &+\frac{\lambda_S}{4}S^4+\frac{\lambda_{HS}}{2}H^{\dagger}H S^2+\frac{\lambda_{\varphi S}}{4} \varphi^2 S^2,
	\end{split}
    \end{equation}
    with the boundary values $m_H^2(\mu_0)=m_{\varphi}^2(\mu_0)\equiv m_{\Sigma}^2$,
    $\lambda_H(\mu_0)=\lambda_{\varphi}(\mu_0)=\lambda_{H\varphi}(\mu_0)\equiv\lambda_{\Sigma}$ and $\lambda_{HS}(\mu_0)
    =\lambda_{\varphi S}(\mu_0)\equiv\lambda_{\Sigma S}$ at the renormalization scale $\mu_0$ featuring the global $\SO(5)$ symmetry. The generic feature of this model framework is that in the case of large hierarchy $v_{\rm{w}}\ll v$, all scalar couplings are parametrically small. This makes the scalar fields appealing inflaton candidates, as we will discuss below.
    
We choose the renormalization scale $\mu_0$ such that the tree-level vacuum expectation value, $v$, is not changed by the one-loop corrections and determine the preferred value of
the vacuum alignment by minimizing the one-loop Coleman--Weinberg potential with respect to $\theta$. Furthermore, we
require that the model provides a pGB Higgs boson with the correct physical mass which we calculate
following Refs.~\cite{Casas:1994us,Gonderinger:2009jp}.

\begin{figure}
\begin{center}
\includegraphics[width=.4\textwidth]{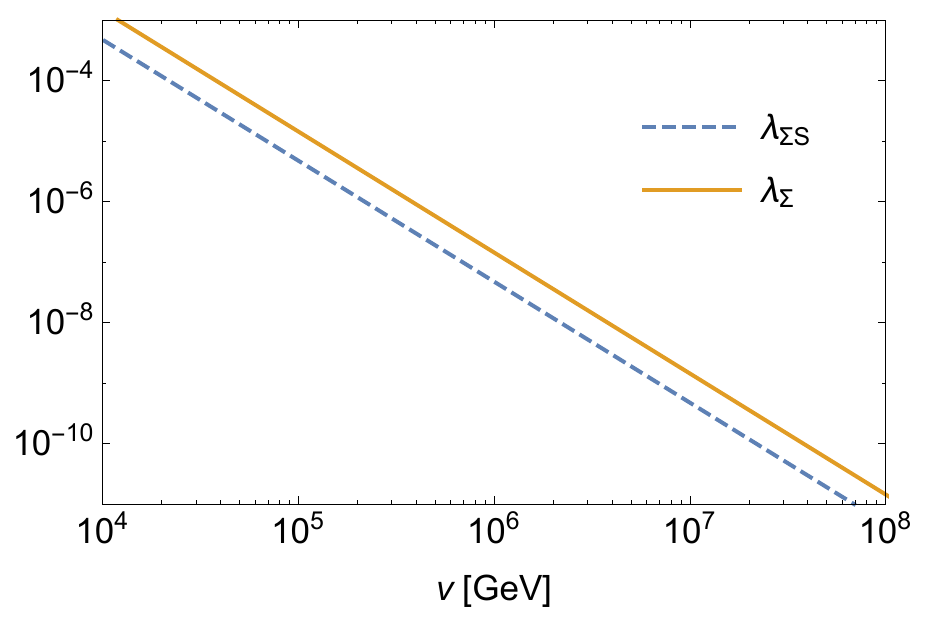}
\caption{Scalar couplings $\lambda_{\Sigma}\equiv\lambda_{H}(\mu_0)$ (solid yellow line), $\lambda_{\Sigma S}\equiv\lambda_{HS}(\mu_0)$ (dashed blue line) as a function of the symmetry breaking scale $v$. The singlet scalar coupling $\lambda_S$ is independent of $v$.}
\label{smallcouplingplot}
\end{center}
\end{figure}

Finally, we choose a benchmark model, for which the two SM-singlet states
have equal masses. This allows us to solve the quartic couplings $\lambda_{\Sigma}$ and $\lambda_{\Sigma S}$ as functions
of the symmetry breaking scale, see Fig.~\ref{smallcouplingplot}, and the resulting renormalization scale is given by
$\mu_0\approx 5.4\times 10^4$~GeV. We assume that the remaining quartic coupling, $\lambda_S$, that is not fixed by the
symmetry-breaking dynamics is of the same order as the other scalar quartic couplings in order not to produce large
hierarchies among these couplings. However, to ensure successful reheating dynamics we take $\lambda_S<\lambda_{\Sigma S}$, as in \cite{Tenkanen:2016twd}.

The fields $H$ and $\varphi$ in this scalar potential are the EW interaction eigenstates.  The relevant physical mass eigenstates will in general be mixtures of the neutral component of $H$ and $\varphi$, such that the Higgs is the mostly Goldstone-like state. The mostly $\varphi$-like eigenstate is heavier. Since we work in the limit where the hierarchy $v_{\rm{w}}\ll v$ is large, $\cos\theta\simeq 1$, and, as a result, the mixing can be neglected for all practical purposes.

In addition, we will include non-minimal couplings to gravity,
\begin{equation}
\label{Vxi}
V_\xi=\xi_H(H^{\dagger}H)R+\frac{1}{2}\xi_{\varphi}\varphi^2\, R+\frac{1}{2}\xi_S S^2\, R,
\end{equation}
again with the boundary condition $\xi_H(\mu_0)=\xi_{\varphi}(\mu_0)\equiv\xi_{\Sigma}$. The presence of such non-minimal couplings is motivated by the analysis of quantum corrections in a curved background,  as they have been shown to generate such terms even if the couplings are initially set to zero \cite{Freedman:1974gs}. We assume again a modest hierarchy $\xi_{\Sigma}<\xi_S$, and assume that other gravitational couplings, such as $\alpha R^2$, are negligible (for their effect on inflationary dynamics, see e.g. \cite{Salvio:2015kka,Calmet:2016fsr}).

Since all scalar couplings are very small, the contributions of ${\cal{O}}(\lambda_i)$,
    $i=H,\varphi,S, H\varphi, HS, \varphi S$, to the $\beta$-functions are negligible and we omit them. In this limit, the only non-zero $\beta$-functions are given by
    \begin{equation}
	\label{eq:}
	\begin{split}
	    16\pi^2\beta_{\xi_H}\simeq &\left(\xi-\frac{1}{6}\right)\left(6y_t^2-\frac{9}{2}g^2-\frac{3}{2}g^{\prime\, 2}\right),\\
	    16\pi^2\beta_{\lambda_H}\simeq &\frac{3}{8}\left(3g^2+2g^2 g^{\prime\,2}+g^{\prime\, 4}\right)-6y_t^4.\\
	\end{split}
    \end{equation}

The important feature is that the couplings of the singlet directions ($\lambda_S$ and $\lambda_{HS}$), which are determined to be small by the the symmetry breaking pattern and successful reheating dynamics, remain small up to the highest scales we consider.

\section{Cosmic inflation}
\label{inflation}

The Jordan frame action, where the non-minimal couplings to gravity are explicit, is
\begin{eqnarray}
S = \int d^4x \sqrt{-g}\bigg(\frac{1}{2}\partial_{\mu}\phi_i\partial^{\mu}\phi^i - \frac{1}{2}M_P^2R - V_\xi - V_0 \bigg) ,
\end{eqnarray}
where $M_P$ is the reduced Planck mass\footnote{Because $\xi_\Sigma v^2 \ll M_{P}^2$, we will neglect the term proportional to $v$ and, as a result, the scale $M_P$ can indeed be identified as the Planck mass to a very good accuracy.}, the sum in the kinetic term goes over $i=H,\varphi, S$, and $V_0$ and $V_\xi$ are given by Eqs. \eqref{V0} and \eqref{Vxi}, respectively.

By making a conformal transformation to the Einstein frame,
\begin{equation}
\hat{g}_{\mu\nu} = \Omega^2 g_{\mu\nu}, \hspace{1cm} \Omega^2\equiv 1+\frac{2V_\xi}{M_P^2R} ,
\end{equation}
the coupling between the scalars and gravity can be made minimal. If we also define
\begin{equation}
\label{h_chi}
\frac{d\chi_i}{d\phi_i} = \sqrt{\frac{\Omega^2+6\xi_i^2\phi_i^2/M_P^2}{\Omega^4}} ,
\end{equation}
for all $i=H,\varphi,S$ and take $H,\varphi = 0$ during inflation by virtue of the assumed hierarchy $\xi_H < \xi_S$ which minimizes the scalar potential in the $S$ direction, the kinetic term of $\chi_S$ becomes canonical. Thus, by taking the Jordan frame potential to be $V_0=\lambda_S S^4/4$, the Einstein frame potential, $U(\chi_S)=\Omega^{-4}V_0$, becomes
\begin{equation}
\label{potential}
U(\chi_S) \simeq \frac{\lambda_S M_P^4}{4\xi_S^2}\left(1+\exp\left(-\frac{2\sqrt{\xi_S}}{\sqrt{6\xi_S+1}}\frac{\chi_S}{M_{P}} \right) \right)^{-2} ,
\end{equation}
at large field values, $S \gg M_P/ \sqrt{\xi_S}$, or $\chi_S \gtrsim M_P$. We see that the potential is exponentially flat and thus sufficient for slow-roll inflation.

Note that if we would have assumed $\varphi,S=0$ and taken the Higgs to drive inflation, we would have arrived at a similar result for $\chi_H$. This is the case of SM Higgs inflation. Because in the SM $\lambda_H \sim 0.1$, the classical potential for $\chi_H$ is sufficient to produce the observed amplitude of the curvature power spectrum only for very large non-minimal coupling values, $\xi_H\sim 10^4$~\cite{Bezrukov:2007ep}.

In our scenario, where the Higgs is a pGB, this hindrance does not arise because now all scalar couplings can be parametrically small, see Fig.~\ref{smallcouplingplot}. Without considerable fine-tuning we cannot, however, allow for $H$ to drive inflation due to potential instability at scales below the inflationary scale
\footnote{See, however, Refs.~\cite{Bezrukov:2014ipa,Fumagalli:2016lls,Enckell:2016xse} for related discussion.}, and therefore we take the singlet $S$ to be the inflaton. For $S$, the potential is stable up to high scales, and the correct amplitude of the curvature power spectrum can be obtained even for $\xi_S < 1$, as we will discuss below.

A similar scenario with a small inflaton self-coupling was recently studied in \cite{Tenkanen:2016twd} in the context of unification of inflationary dynamics and dark matter production. By following similar steps, we present here the main results for the three inflationary observables $\mathcal{P}_\mathcal{R}, n_s$ and $r$, and discuss what matching them to the observed values of $\mathcal{P}_\mathcal{R}$ and $n_s$ require in terms of the model parameters.

The usual slow-roll parameters are defined as
\begin{eqnarray}
\epsilon &\equiv& \frac{1}{2}M_P^2 \left(\frac{dU(\chi_S)/d\chi_S}{U(\chi_S)}\right)^2 ,
\\ \nonumber
\eta &\equiv& M_P^2 \frac{d^2U(\chi_S)/d\chi_S^2}{U(\chi_S)} ,
\end{eqnarray}
and the COBE satellite normalization \cite{Lyth:1998xn} requires
\begin{equation}
\label{cobe}
\frac{U(\chi_S)}{\epsilon} = 0.027^4M _P^4 
\end{equation}
to give the correct amplitude for the curvature power spectrum, $P_\mathcal{R} \simeq 2.2\times 10^{-9}$ \cite{Ade:2015lrj}. The requirement in Eq.~(\ref{cobe}) can be expressed in terms of inflationary e-folds, $N=\ln(a_{\rm end}/a)$, as
\begin{equation}
\label{xi}
\frac{2\lambda_S N^2}{6\xi_S^2+\xi_S}\simeq 0.027^4 ,
\end{equation}
which gives an approximate estimate for the required value of the non-minimal coupling $\xi_S$ in terms of $\lambda_S$ and $N$. We see that in the limit $\lambda_S\rightarrow 1$ the non-minimal coupling indeed has to be very large, $\xi= \mathcal{O}(10^4)$, to be consistent with observations. 

\begin{figure}
\begin{center}
\includegraphics[width=.4\textwidth]{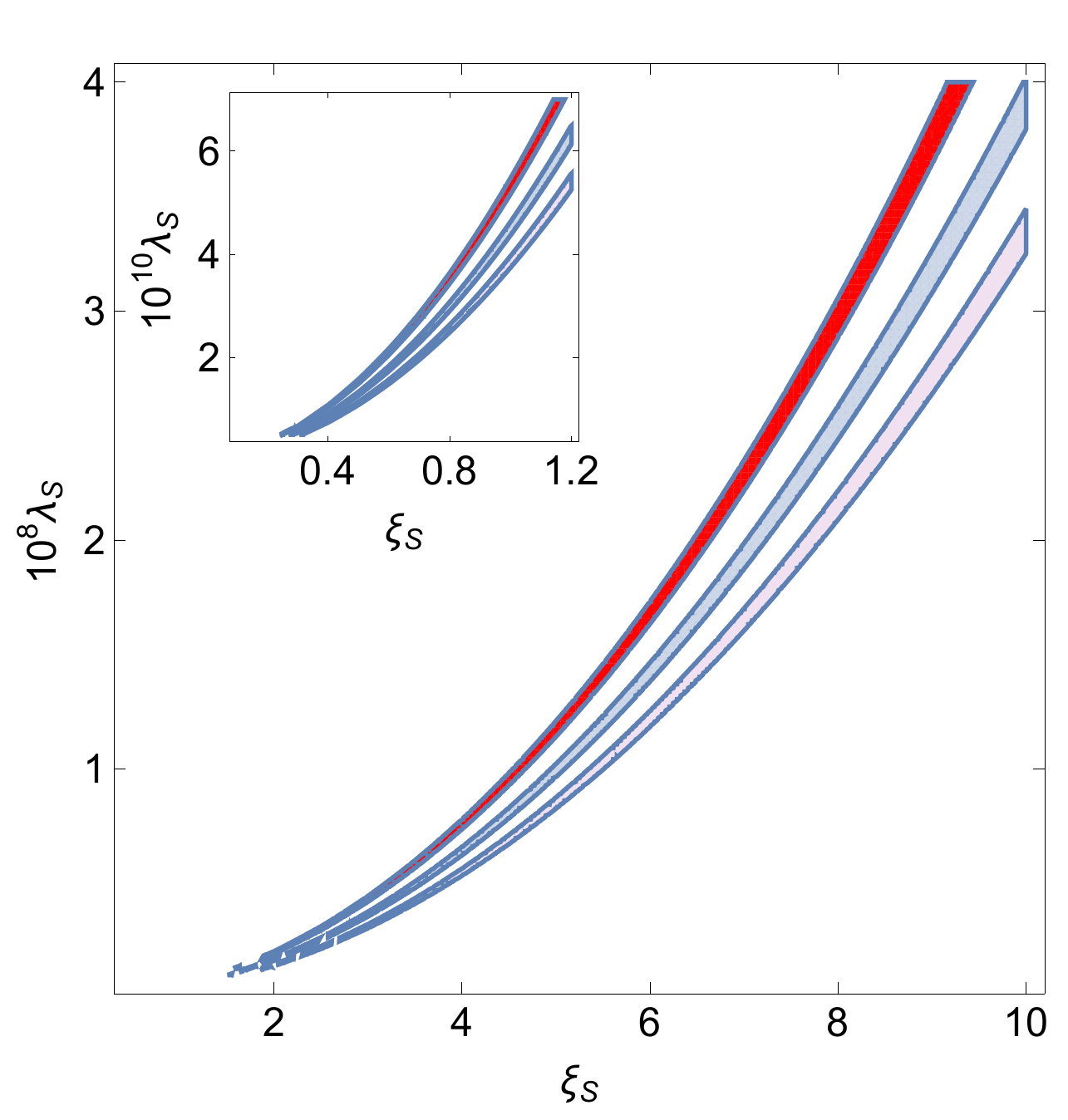}
\caption{The curvature power spectrum $\mathcal{P}_\mathcal{R} = (2.139 \pm 0.063)\times 10^{-9}$ in the $(\xi_S,\lambda_S)$-space for $N=55$ (red), $N=60$ (blue), and $N=65$ (light purple).}
\label{PR}
\end{center}
\end{figure}

In our analysis we, however, compute the curvature power spectrum numerically from $U(\chi_S)=\Omega(S(\chi_S))^{-4}V_0(S(\chi_S))$, where $S(\chi_S)$ is given by  Eq.~\eqref{h_chi}. The results are presented in Fig. \ref{PR} in terms of $\xi_S$ and $\lambda_S$ for representative values of $N$. We note that in this scenario, already the tree-level estimates for inflationary observables are very accurate due to negligible quantum corrections to $\lambda_S$, as discussed above.

For the spectral index $n_s-1 \simeq -6\epsilon+2\eta$ and tensor-to-scalar ratio $r\simeq 16\epsilon$, we obtain the results
\begin{equation}
\label{nsr}
n_s \simeq 0.9678, \hspace{.5 cm} 0.0030 < r < 0.0078,
\end{equation}
for $0.1<\xi_S<10$ and $N\simeq 60$. For this range, we have $\xi_H\sim 0.1$ to maintain consistency with our
initial assumption $\xi_H< \xi_S$. The results match very well to the observed value $n_s = 0.9677 \pm 0.0060$ and to the upper limit $r\leq 0.11$~\cite{Ade:2015lrj}. 

In \cite{Tenkanen:2016twd} it was shown that in a scenario similar to ours the reheating temperature is\footnote{We have assumed that thermal corrections to $m_H$ are negligible during reheating, and checked that extending the calculations of \cite{Tenkanen:2016twd} to the range $0.1\leq\xi\leq 1$ does not change this result.}
\begin{equation}
T_{\rm RH} \simeq 0.002\lambda_{HS}^2\lambda_S^{-3/4}M_P ,
\end{equation}
which for $\lambda_{HS}\gtrsim \lambda_S \gtrsim 10^{-10}$ is well above the scales related to the Big Bang Nucleosynthesis. We therefore conclude that our scenario is in accord with the most important cosmological observations. While at the classical level there are no observable consequences that would distinguish the model from other models of the same type, the inclusion of loop corrections to specific model setups can have an effect on inflationary observables, as recently demonstrated in e.g. \cite{Tenkanen:2016idg}, which would allow distinguishing our model from others.

Finally, we note that the potential problems of the SM Higgs inflation associated with unitarity violation during
inflation~\cite{Burgess:2009ea,Barbon:2009ya,Barvinsky:2009ii,Bezrukov:2010jz,Burgess:2010zq,Hertzberg:2010dc}
do not arise at all in the models of the type we study
because the scale of perturbative unitarity breaking is always higher than the scale of inflation~\cite{Calmet:2013hia,Kahlhoefer:2015jma}.

\section{Higgs vacuum stability}
\label{vacuumstability}
In the SM the vacuum can be metastable in the Higgs direction \cite{Degrassi:2012ry}, although this result is subject to
some uncertainty considering the relation between the mass measured by the experiments and
the pole mass entering the theoretical computation~\cite{Bezrukov:2014ipa,Iacobellis:2016eof}. In our model the scalar potential is
stable in all directions except the one corresponding to the Higgs, and the situation is essentially
similar to the SM. However, since in our case the Higgs self-coupling is smaller than in the SM, the
possible problem is actually a bit worse than in the SM.

\begin{figure}
	 \begin{center}
	     \includegraphics[width=0.4\textwidth]{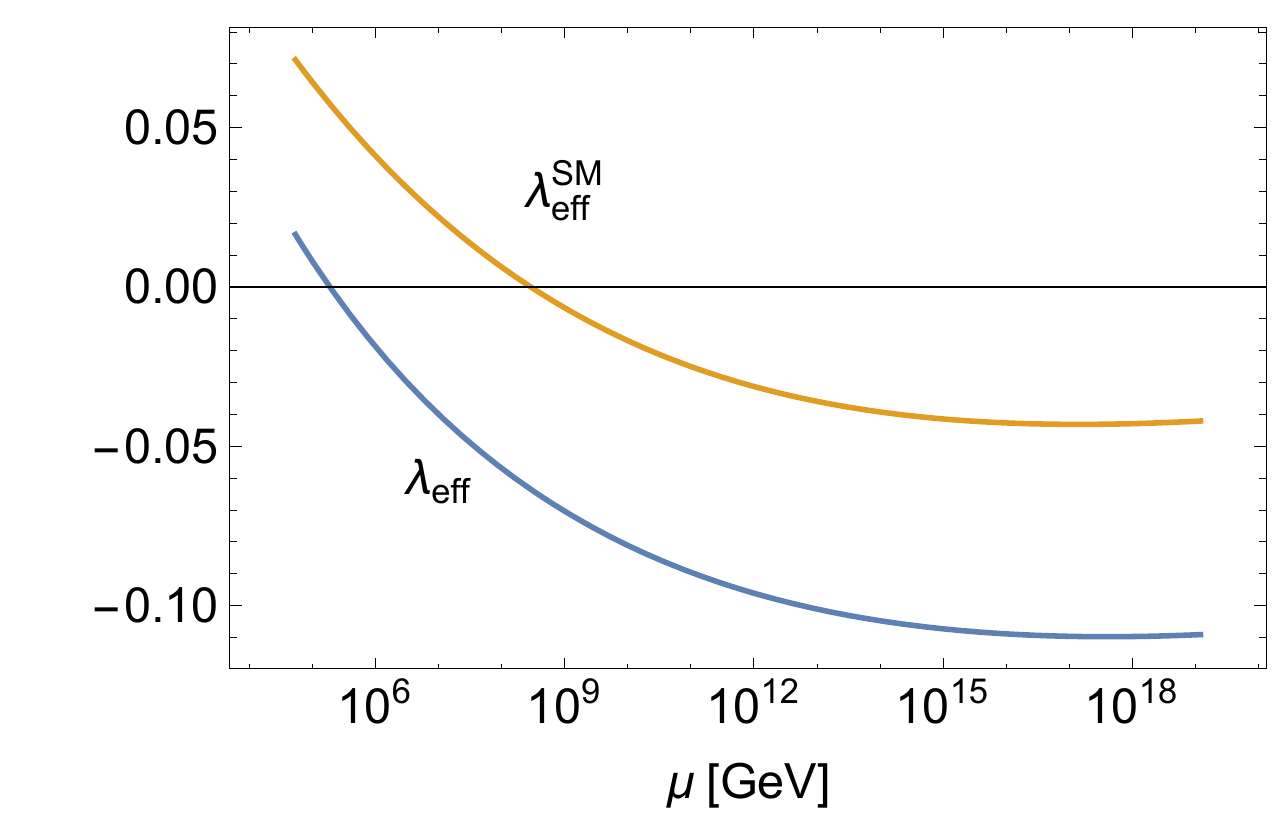}
	 \end{center}
	 \caption{One-loop running of $\lambda_{\mathrm{eff}}$ compared to one-loop running in the SM.}
	 \label{fig:1loopRG}
    \end{figure}

To analyse the situation in more detail, we consider the one-loop RG-improved effective potential in the Higgs direction,
    \begin{equation}
	\label{eq:}
	V_{\mathrm{eff}}(h)=\frac{\lambda_{\mathrm{eff}}(h)}{4}h^4,
    \end{equation}
    where the effective self-coupling is determined as explained in~\cite{Degrassi:2012ry}. The running of $\lambda_{\mathrm{eff}}$ with the initial condition $\lambda_H(5.4\times 10^4\,\mathrm{GeV})=10^{-9}$ is shown in Fig.~\ref{fig:1loopRG}.
As in the SM, the scale at which $\lambda_{\mathrm{eff}}=0$ is somewhat increased when higher-order corrections are taken
into account.

As observed in~\cite{Herranen:2014cua,Herranen:2015ima}, the value we have assumed for the non-minimal gravity coupling of the Higgs field, $\xi_H=\mathcal{O}(0.1)$, lies exactly in the region where the vacuum is expected to be stable against large fluctuations both during and immediately after inflation.
In particular, following the treatment in Ref. \cite{Herranen:2014cua}, we obtain a lower limit $\xi_H(H^\ast)\geq 0.05$ for stability during inflation,
and the upper limit from gravitational particle production after inflation, $\xi_H(H^\ast)<3/8$, is unchanged
with respect to~\cite{Herranen:2015ima}. 

To study whether the vacuum remains metastable at field values less than the inflationary scale, we apply the standard flat-space analysis, see e.g.~\cite{Isidori:2001bm}. The condition for metastability in this case is~\cite{Antipin:2013sga}
\begin{equation}
    \label{eq:metastability}
    \lambda_{\mathrm{eff}}(h)>-\frac{2\pi^2}{3\log[h T_{\mathrm{U}}\mathrm{e}^{\gamma_{\mathrm{E}}}/2]},
\end{equation}
where $T_{\mathrm{U}}$ is the age of the universe, and $\gamma_{\mathrm{E}}$ is the Euler–Mascheroni constant.
Higher-order corrections are expected to raise the scale where $\lambda_{\mathrm{eff}}$ becomes negative by at least two orders of magnitude. Taking this into account, we conclude on basis of Eq.~\eqref{eq:metastability}, that the vacuum remains in the metastabe region under fluctuations below the inflationary scale.

\section{Conclusions and Outlook}
\label{checkout}

In this paper, we have considered a scenario which attempts to address some of the shortcomings of the simplest Higgs
inflation scenario. Concretely, we considered a scalar sector based on the idea of the Higgs boson as a pseudo-Goldstone boson. When
realized in terms of elementary scalars, the simplest model of this type requires a singlet scalar to be introduced
in order to break the electroweak symmetry of the vacuum. We identified this singlet field as an inflaton and studied the
resulting model in light of current precision cosmology observations.

Our results show that within this framework one can avoid the introduction of very large
non-minimal couplings to gravity, and still maintain the successful predictions from inflation.
From the model building point of view, our model shows
how the inflationary dynamics can be fully embedded in a coherent particle physics setting: the electroweak symmetry
breaking as a whole arises from the dynamics of the inflaton. This should pave the way for similar investigations
in other model setups based on extensions of the electroweak sector of the Standard Model.

\section*{Acknowledgements}
We thank S. Nurmi for disucssions. This work has been supported by the Academy of Finland, grant\# 267842. The CP$^3$-Origins centre is partially funded by the Danish National Research Foundation, grant number DNRF90. T.T. is supported by the Research Foundation of the University of Helsinki and the U.K. Science and Technology Facilities Council grant ST/J001546/1. T.A. acknowledges partial funding from a Villum foundation grant.

\bibliography{Inflation.bib}

\end{document}